# Clustering based magnetic assays for SARS-CoV-2 detection with scFv-functionalized magnetic nanoparticles

F. T. Wolgast[1*], N. Lehmler[2], S. Westerhoff[2], S. Demiral[2,3], J. Lohmann[3], R. Amin[1], R. Sack[1], A. Lak[1], M. Schilling[1], M. Schubert[2,3], T. Viereck[1]

[1] Institute for Electrical Measurement Science and Fundamental Electrical Engineering and the Laboratory for Emerging Nanometrology (LENA), TU Braunschweig, 38106 Brunswick, Germany

[2] Institute for Biochemistry, Biotechnology and Bioinformatics, Department of Biotechnology, TU Braunschweig, 38106 Brunswick, Germany

[3] Helmholtz Centre for Infection Research, VLP-Based Technologies, 38124 Brunswick, Germany

* Corresponding Author, f.wolgast@tu-braunschweig.de

## Abstract

Magnetic nanoparticles (MNP) can be used in magnetic immunoassays (MIA) by functionalizing them with antibodies. In homogeneous assays that follow a "mix-and-measure" approach, it is then sufficient to add the sample in question and evaluate the response of the MNPs using a magnetic measurement method. Magnetic Particle Spectroscopy (MPS) is a fast and sensitive method for this purpose, capable of determining the binding state by measuring changes in Brownian relaxation. For further analysis, alternating current susceptometry (ACS) is considered, although it is significantly more time-consuming and therefore no option for applications interested in point of care detection. In this study, we aim to further improve the approach of MIA evaluated by MPS. To this end, we use self-synthesized scFv fragments instead of whole IgG antibodies during functionalization in order to keep the size of the BNF-Dextran MNP with 80 nm nominal diameter as small as possible, which allows for greater relative size changes upon binding to an analyte. Virus-like particles (VLP) and the N protein of SARS-CoV-2, which were also produced in-house, are used as analytes. Due to multiple binding sites, these form cluster structures, which, in the case of VLP, were investigated in greater depth ACS to assess their field dependence. In addition, the sensitivity of the assays was analyzed as a function of MNP concentration, and the detection limit for both analytes was estimated. We found that, using scFv-functionalized MNPs, we were able to detect low concentrations of just 490 fM of SARS-CoV-2 VLP and 1.7 nM of the N protein. However, the slightly above-proportional increase in sensitivity as the MNP concentration decreases does not automatically imply a better detection limit.

## Introduction

Research into the principle of magnetic assays for the detection of pathogens and proteins has been ongoing for years and has received increased attention in the wake of the COVID-19 pandemic. Among the various methods available for analysing such assays, Magnetic Particle Spectroscopy (MPS) has established itself as a promising approach. MPS has its origins in the development of magnetic particle imaging (MPI) and served as a platform for the evaluation of magnetic nanoparticles (MNP) [Biederer2009]. However, the characteristic magnetic properties measured in this process can also be used for biosensor analysis if the MNP are functionalized accordingly beforehand [Krause2007, Nikitin2007]. Thus, magnetic immunoassays (MIA) were used to successfully detect Staphylococcal Toxins [Orlov2013], Botulinum Neurotoxin [Orlov2016], Cholera toxin [Achtsnicht2019], plant viruses [Rettcher2015] or the of Influenza A virus subtype H1N1 [Wu2020]. The detection of SARS-CoV-2 was also successfully demonstrated. Thereby the target substances included not only the Nucleocapsid and Spike proteins as well as RNA, but also mimic viruses made from polystyrene beads [Wu2021, Zhong2021, Rösch2021]. In addition, the detection of the corresponding antibodies has been demonstrated [Vogel2022]. The overall focus is on developing rapid, homogeneous and wash-free assays as an alternative to the currently established testing methods in the field of point-of-care (POC) diagnostics. While reverse transcription-polymerase chain reaction (RT-PCR) is a highly sensitive method for detecting DNA/RNA, it requires expensive laboratory equipment and reagents, involves a complex workflow, and is susceptible to contamination, making it unattractive for these applications. In addition, there are lateral flow devices (LFD) for antigen detection, which are inexpensive and can be used independently, but have limited specificity and sensitivity and are available only for certain diseases [Everitt2021, Cassedy2021, Mak2020].

Given this background, homogeneous MIA exhibit several properties that are of interest for the point-of-care testing. First, they offer short analysis times, since as with LFD no filtering, washing or purification steps are required. At the same time, they are easy to handle, as the MNP simply need to be mixed with the analyte and can be measured directly after an incubation period. Direct magnetic measurement also makes the method robust against the optical background of the sample and allows for the analysis of turbid or opaque media [Wu2022, Cao2020]. However, this does not necessarily translate into improved analytical performance. In particular, with regard to the limit of detection, MIA has not yet demonstrated any general advantages over established methods and is therefore the subject of current research. Current approaches include improved measurement devices [Chugh2021, Wolgast2024], expanded measurement methods [Vogel2022, Liu2023], and the use of amplification circuits in the field of DNA/RNA assays [Rösch2024, Sack2025]. Another key factor are the MNP used [Chowdhury2023] as well as their functionalization [Kahmann2024].

In this paper, we further explore the functionalization of MNP by using only the scFv part instead of the usual full-length IgG. We also expand the scope of SARS-CoV-2 detection by including virus-like-particles (VLP) [Korn2020]. Moreover, we examine the clustering and its effects on MPS measurements frequently observed in MIA, and investigate the influence of different MNP concentrations. Finally, an estimate of the current limit of detection (LOD) for SARS-CoV-2 VLP and their Nucleocapsid proteins is provided.

# Methods and Materials

The following sections provide an overview of the experimental procedures used throughout this study. This includes the magnetic measurement techniques, the assay components, and the preparation of the respective sample series.

## Physical Principles of homogeneous Magnetic Assays

The analysis of homogeneous MIA is based on the hydrodynamic size change of the particles. This occurs when an MNP functionalized with covalently bound antibodies encounters the antigen that specifically matches those antibodies. For spherical particles, an increase in the hydrodynamic diameter $d_h$ results in a slower Brownian motion [Rauwerdink2010]. The dynamic of this mechanism is described by the zero-field Brownian relaxation time constant

$$\tau_{B0} = \frac{\pi \eta d_h^3}{2 k_B T}, \tag{1}$$

which depends, in addition to the thermal energy $k_B T$, where T is the temperature and $k_B$ denotes the Boltzmann constant, on the viscosity η. Strictly this formula is only valid for small-field excitations. For larger field excitations exceeding the approximately linear part of the magnetization, such as those in the MPS, the zero-field constant $\tau_{B0}$ must be replaced by the field-dependent one, which is given by [Ludwig2013]

$$\tau_{BH} = \frac{\tau_{B0}}{\sqrt{1 + 0.126 \xi^{1.72}}} \tag{2}$$

with $\xi = mB/(k_B T)$ including the magnetic energy mB, which is determined by the magnetic moment m of an MNP and the magnetic flux density B of the excitation. In addition, there is another relaxation mechanism known as Neél relaxation. For our homogeneous MIA, it is required that MNP are chosen that this mechanism is thermally blocked, and thus Brownian relaxation dominates. Since the particle relaxation dynamics directly influence the magnetization response, changes in particle motion can be detected using magnetic measurement methods like alternating current susceptometry (ACS) or MPS.

## Magnetic Measurement Methods

ACS uses small sinusoidal magnetic fields which periodically excite the nanoparticles in the linear domain while sweeping the excitation frequency to measure the magnetic susceptibility $\chi$. Although this method is sensitive and provides information about the distribution of relaxation times of MNP, which in turn allows conclusions to be drawn about the particle size distribution [Dieckhoff2014, Ludwig2017], the measurement takes a long time to cover the full frequency spectrum. Moreover, the fundamental is affected by the para- or diamagnetic contributions of the surrounding environment, for instance the water in the ferrofluid or the coil bobbin. In contrast, although MPS uses a similar setup, higher excitation amplitudes at only one frequency with high spectral purity are employed to exploit the nonlinear magnetization response of the MNP. For analysis, the time-domain signal is recorded and transformed into the frequency domain using a Fourier transform. The resulting higher harmonics are specific to the MNP and can be used to distinguish between different samples. Therefore, MPS promises the advantage of a fast, simple, cost effective and sensitive evaluation platform. In this work, ACS and MPS measurements are performed using the RMF system in ACS mode and the immunoMPS setup described in [Dieckhoff2011] and [Wolgast2024], respectively. The MPS system measures a sample volume of 60 µL at an excitation frequency of 590 Hz and a field strength of 15 mT/$\mu_0$, while the ACS measurements are performed with 150 µL and over a frequency range from 1 Hz up to 3 kHz at 0.2 mT/$\mu_0$. In our experiments, each sample is measured five times in MPS measurements, whereas in the ACS only one run is performed due to the duration. For analysis the harmonic ratio $HR_{53} = H_5 / H_3$ is typically used in the MPS data evaluation, since, unlike individual harmonics such as the fifth ($H_5$) and the third ($H_3$), this is a concentration-independent metric. In ACS measurements, the imaginary part of the susceptibility $\chi''$ is primarily considered because it allows conclusions to be drawn about the hydrodynamic size distributions of the MNP. The normalized representation $\chi''_{norm}$ is particularly well-suited for this purpose, as it eliminates variations in concentration.

## Magnetic Nanoparticles

Our assay platform relies on BNF-Dextran particles (micromod GmbH, Rostock, Germany) exhibiting a hydrodynamic diameter of 80 nm (BNF-D80) and provided with a streptavidin coating. They have a particle mass concentration of $\beta_{MNP}$ = 10 mg/ml and a particle density of $C_{MNP} = 1.2 \cdot 10^{13}$ particles/ml, respectively, which corresponds to a molar concentration of $c_{MNP}$ = 19.92 nM. Furthermore, the nominal iron concentration is $\beta_{Fe}$ = 5.5 mg/ml, and the biotin binding capacity is specified as $q_B$ > 100 pmol/mg(Fe). This guarantees a minimum of $N_{BBS}$ > 27.6 biotin binding sites (BBS) per MNP. Based on the batch-specific certificate of analysis (available upon request), $N_{BBS}$ can be more accurately determined, typically ranging from 50 to 100 BBS/MNP. This gives an initial estimate of how many molecules can be used to functionalize the particles.

## Antibodies

To this end, the following four types of antibodies are considered: one IgG and three scFv antibodies. In all cases, the antibodies are produced in-house and biotinylated so that they can be bound to the surface of the MNP, thereby enabling specific binding to the target substances. The IgG (STE90-C11-IgG) is a monoclonal antibody targeting the SARS-CoV-2 Spike (S) protein [Bertoglio2021a]. It has a molar mass of 143.532 kDa and was randomly biotinylated after production using EZ-Link Sulfo NHS LC

Biotin kit (Thermo Fisher Scientific) according to the manufacture's manual for 4-6 biotin per IgG. Successful biotinylation was confirmed in ELISA on immobilized RBD protein. The corresponding scFv fragment (STE90-C11-scFv) has a calculated weight of 30.968 kD. In addition, two scFv fragments (SH2424-B1-scFv with a molecular mass of 31.768 kDa & SH2424-H9-scFv with a molecular mass of 32.184 kDa) targeting the Nucleocapsid (N) protein of SARS-CoV-2 were generated in house by phage display [Nur2023]. All antibodies and scFv were produced using the Expi293F expression system described before [Bertoglio2021b]. The scFv include a His tag for purification as well as an Avidin Tag for site specific biotinylation *in vivo* by co-expression of the birA biotin ligase [Fairhead2015]. The biotinylation ratio determined in Shift assay [Fairhead2015] and analysed by ImageJ are ~53 % for STE90-C11, ~73 % for SH2424-B1 and ~26 % for SH2424-H9 respectively.

Kinetic characterization was performed via Bio-Layer Interferometry (BLI) using a Gator Prime (Gator Bio, Palo Alto, CA, USA) at 30 °C under constant shaking at 1000 rpm. Each step was conducted in kinetic buffer (TBS, 0.1 % PVP, 0.02 % Tween-20). Streptavidin biosensors (Streptavidin (SA) Probes, PN 160002, Gator Bio) were loaded with scFv at a concentration of 62.5 nM until a shift of 1 nm was reached. Association of the antigens was recorded for 180 s using a 1:3 serial dilution ranging from 20 nM to 0.08 nM (N protein) or 40 nM to 0.5 nM (RBD25), followed by dissociation in kinetic buffer for 600 s. To account for systematic drift and non-specific binding, a double-referencing strategy was applied. For each run, two control sensors were included: a buffer-reference sensor (scFv-loaded, but immersed in buffer during the association phase) to correct for baseline drift, and a matrix-reference sensor (unloaded sensor immersed in the highest antigen concentration) to monitor non-specific binding to the sensor surface. Data analysis was conducted using the Gator Analysis Software (Version 2.19.8.0424, Gator Bio). Sensorgrams were aligned to the baseline, and kinetic parameters association rate constant ($k_{on}$), dissociation rate constant ($k_{off}$) and equilibrium dissociation constant ($K_D$) were calculated globally using a 1:1 binding model with $R_{max}$ unlinked. This specific assay orientation was chosen to closely mimic the surface architecture of scFv-coated magnetic nanoparticles, thereby ensuring comparability of the kinetic data.

## Sandwich-ELISA

In a high-binding 96-well plate (Costar, Bio-Rad Laboratories GmbH, USA) 200 ng/well of the RBD-binding human IgG STE90-C11 was immobilized by incubation for 1 h at room temperature (RT). Followed by blocking of the wells with 2 % *(w/v)* BSA in PBS over night at 4 °C. The 96-well plate was washed three times with 170 µL of PBS-T (0.05 % *v/v*) by hand. Afterwards the VLP were diluted on the well plate according to the respective dilution series and incubated for 1.5 h at RT. Washing was performed again. The site specific biotinylated scFv-fragment STE90-C11-scFv-Avitag was added in a concentration of 200 ng/well. After incubation for 1 h at RT, washing was performed and HRP-conjugated streptavidin was added (RABHRP3, Sigma-Aldrich, St. Louis, MI, USA, diluted 1:500) for 1 h at RT. Washing was performed, followed by the addition of 100 µL TMB/E solution (ES001, Merck Millipore, Burlington, VT, USA). After 45 minutes, 1 N $H_2SO_4$ was added to stop the reaction. Finally, the absorbance at 450 nm and at 620 nm as reference was measured, using a microplate reader (Spark® Multimode Microplate Reader, Tecan Group, Männedorf, Switzerland).

## Virus-Like-Particles and Proteins

The substances to be detected included SARS-CoV-2 VLP (mean diameter of ~140 nm) and its N protein (dimer, calculated mass at 91.706 kDa). In addition, negative controls were needed to verify the specificity of the assays. For this purpose, VLP of the Hanta virus (Dobrava variant, diameter of ~120 nm) and the SARS-CoV-2 S1-His protein (monomer calculated mass at 76.5 kDa) were selected because they are of comparable size to their positive counterparts.

The soluble antigens were produced in the plasmid-based insect cell expression system as described before [Korn2020]. For purification HisTrap excel column (Cytiva) Äkta on Go and Pure systems (Cytiva) was used followed by preparative Size Exclusion Chromatography on 16/600 Superdex 200 kDa pg column (Cytiva) where also dimerization of N protein was confirmed.

For SARS-CoV-2 Wuhan VLP stabilized full-length Spike, Membrane and Envelope protein and (if indicated) N protein were co-expressed in insect cells and the resulting VLP were concentrated in a 30% Sucrose cushion centrifugation or PEG concentration as described before [Korn2020]. For Dobrava Hanta VLP, Gn and Gc were co-expressed on separate OpiE2 expression vectors. Due to C-terminal fusion of M-Protein (SARS-CoV-2) and Gc-Protein (Hanta) to mcherry their concentration could be determined in Flow Virometry using the Cytoflex Nano (Beckmann Coulter).

## Functionalization and Washing Procedures

The BNF-D80 MNP are functionalized by mixing the streptavidin-coated MNP with the desired, pre-biotinylated antibody. This is followed by an incubation step for 1 h at 25 °C, during which the samples are gently shaken at 600 rpm. They are then stored at 4 °C. This is followed by a two-step washing procedure, as required by the experiment. For this, the functionalized BNF-D80 particles are centrifuged twice for 1 h at 2400 rcf. The resulting supernatant is discarded and replaced with fresh PBS containing 0.05 % Tween20 (PBST). Afterwards the sample is shaken to resuspend the sediments settled at the bottom. To compensate for particle loss, less buffer is added in the final step than is removed as supernatant, assuming an empirical particle loss of 16%.

## Antibody Functionalization Series

A concentration series is performed to determine when and whether the BNF-D80 particles are fully functionalized with antibodies. For the STE90-C11-IgG series, a BNF-D80 concentration of 0.4 nM was prepared for each sample in a volume of 240 µL. The experiment examines changes in the occupancy of the MNP's BBS from 0 % to 100 % in 10 % increments. For the STE90-C11-scFv, the SH2424-B1-scFv and the SH2424-H9-scFv series each sample was prepared with 0.2 nM BNF-D80 in a volume of 220 µL. A wider antibody concentration range is covered from roughly 10 % up to 1000 % of the possible BBS per MNP. A 0 % sample is included as control. The concentrations series was performed logarithmically with 4 samples per decade. To minimize pipetting errors, an iterative dilution method was used. Following mixing with MNP, samples were incubated without the need for further washing steps.

## Cluster Formation Series

The potential effects of clustering in our MIA on the magnetic measurement signals of ACS and MPS are studied with VLP induced clusters. For this purpose, 1200 µL of BNF-D80 particles at a concentration of 0.5 nM are functionalized with STE90-C11-scFv following the general procedure described above. The functionalization was carried out using a fivefold molar excess of scFv relative to the available BBS, which were determined to be $N_{BBS}$ = 107.1 BBS/MNP for this batch. The samples were subsequently incubated and washed. During the washing steps, 1150 µL of supernatant was removed both times and replaced with fresh PBST. In the final step, only 958 µL of PBST was added to compensate for the particle loss. Following the washing process, the functionalized MNP are used for VLP assays with a binary concentration series ranging from roughly 16 pM to 0.25 pM for both SARS-CoV-2 and Hanta VLP. Each assay sample contains 0.2 nM MNP and has a volume of 160 µL. Before measurement, the mixture was incubated for 1 h at 25 °C and 600 rpm.

## Assay Response Series

To investigate the concentration dependence of MIA in combination with MPS and to determine the detection limits of the method, additional samples are required. Therefore, the BNF-D80 particles are functionalized as described for the Clustering Characterization Series. To obtain a sufficient number of samples, 1200 µL of the functionalized MNP suspension with a MNP concentration of 1.5 nM is prepared. The suspension is then incubated separately with STE90-C11-scFv and SH2424-B1-scFv, each at a threefold molar excess relative to the MNP binding-site density of $N_{BBS}$ = 107.1 BBS/MNP. The corresponding washing steps are performed after the incubation. The SH2424-H9-scFv is not considered further, as the previous data suggest only slight differences in binding affinity compared to the SH2424-B1-scFv, while exhibiting a lower biotinylation ratio. The assays were then performed by mixing the functionalized MNP with concentrations of 0.2 nM, 0.1 nM, 0.05 nM, and 0.025 nM with SARS-CoV-2 VLP and N proteins, respectively. While 160 µL was used for the 0.2 nM MNP samples, the sample volume for the remaining MNP concentration levels was 80 µL. The two analytes were tested over a logarithmically spaced concentration range with 5 concentrations per decade using the iterative dilution method. For the SARS-CoV-2 VLP, this covers a range from 13.07 pM down to 210 fM, while for the N protein, the range is from 5.1 nM down to 81 pM. Samples with no target analyte and the not binding Hanta VLP and S1-His proteins are prepared as negative controls at high concentrations, respectively. All the samples are measured after incubation for at least one hour at 25 °C and 600 rpm.

## Detection Limit Series

The experiments on the limit of detection (LOD) build on those from the concentration study and simply expand upon them by including two additional samples for the most relevant MNP concentrations of 0.05 nM and 0.025 nM for both the SARS-CoV-2 VLP and the N proteins. This means that a total of three independent samples are available for these nanoparticle concentrations.

## Methodology for Determining the Limit of Detection

The detection limit from the MPS measurements is derived from the mean blank value $\overline{HR}_{53,blank}$ and three times the estimated standard deviation of an individual measurement, $\hat{\sigma}_{53,blank}$, of the harmonic ratio $HR_{53}$. These quantities define the detection threshold according to

$$HR_{53,thr} = \overline{HR}_{53,blank} + 3\hat{\sigma}_{53,blank}. \quad (3)$$

This threshold does not directly represent the detection limit. Instead, it defines a threshold line in the $HR_{53}$ calibration plot. The LOD is obtained from the intersection of this line with the linear fit of $HR_{53}$ as a function of analyte concentration. The estimation of $\hat{\sigma}_{53,blank}$ requires separating the measurement system-related technical and between-preparation contributions to the total measurement variance. Each of the three independent preparations is measured five times and characterized by its preparation mean $\overline{HR}_{53,i}$ and the corresponding sample variance $s^2_{53,i}$. The overall mean $\overline{HR}_{53}$ is then calculated from the three preparation means. However, the standard deviation calculated from the preparation means does not provide an estimate of the variance of individual measurements, since each preparation mean already represents the average of five observations. The resulting standard deviation therefore reflects the variability between preparations, with only a reduced technical variability contribution. To determine the variance relevant to individual measurements, the technical variance component and the variance component between preparations were therefore estimated separately using a variance component analysis. It yields estimates of the technical and between-preparation variance components $\hat{\sigma}^2_{tech}$ and $\hat{\sigma}^2_{prep}$, respectively. The estimated variance of an individual measurement is then obtained as

$$\hat{\sigma}^2_{single} = \hat{\sigma}^2_{prep} + \hat{\sigma}^2_{tech}. \quad (4)$$

This variance represents the total measurement variability and is therefore used to determine the standard deviation required for the LOD threshold. Details of the variance-component estimation are provided in the Supporting Information.

## Results and Discussion

After confirming the high affinity and functional integrity of the antibodies (biotinylated scFv) by BLI, the magnetic assays were systematically characterized with respect to the MNP functionalization, target-induced clustering, nanoparticle dependent assay sensitivity, and the achievable detection limit.

### Characterization of scFv binding kinetics using Bio-Layer Interferometry

To evaluate the binding affinities and kinetic rate constants of the scFv SH2424-B1, SH2424-H9 and STE90-C11 towards their corresponding antigens, measurements were performed via BLI. To resemble the scFv-coated nanoparticles Streptavidin biosensors were loaded with the biotinylated scFv and the antigens were associated. All binding curves were fitted using a 1:1 binding model ($R^2 > 98$) (Figure 1), yielding the kinetic parameters summarized in Table 1. SH2424-B1 and SH2424-H9 show comparable association and dissociation rate constants, resulting in similar, subnanomolar apparent $K_D$ values. STE90-C11 displayed a slower association rate and a faster dissociation rate, yielding a $K_D$ of 2.7 nM, in agreement with previously published data [Bertoglio2021a]. Rapid association kinetics promotes efficient antigen capture, whereas low dissociation rates reflect enhanced complex stability.

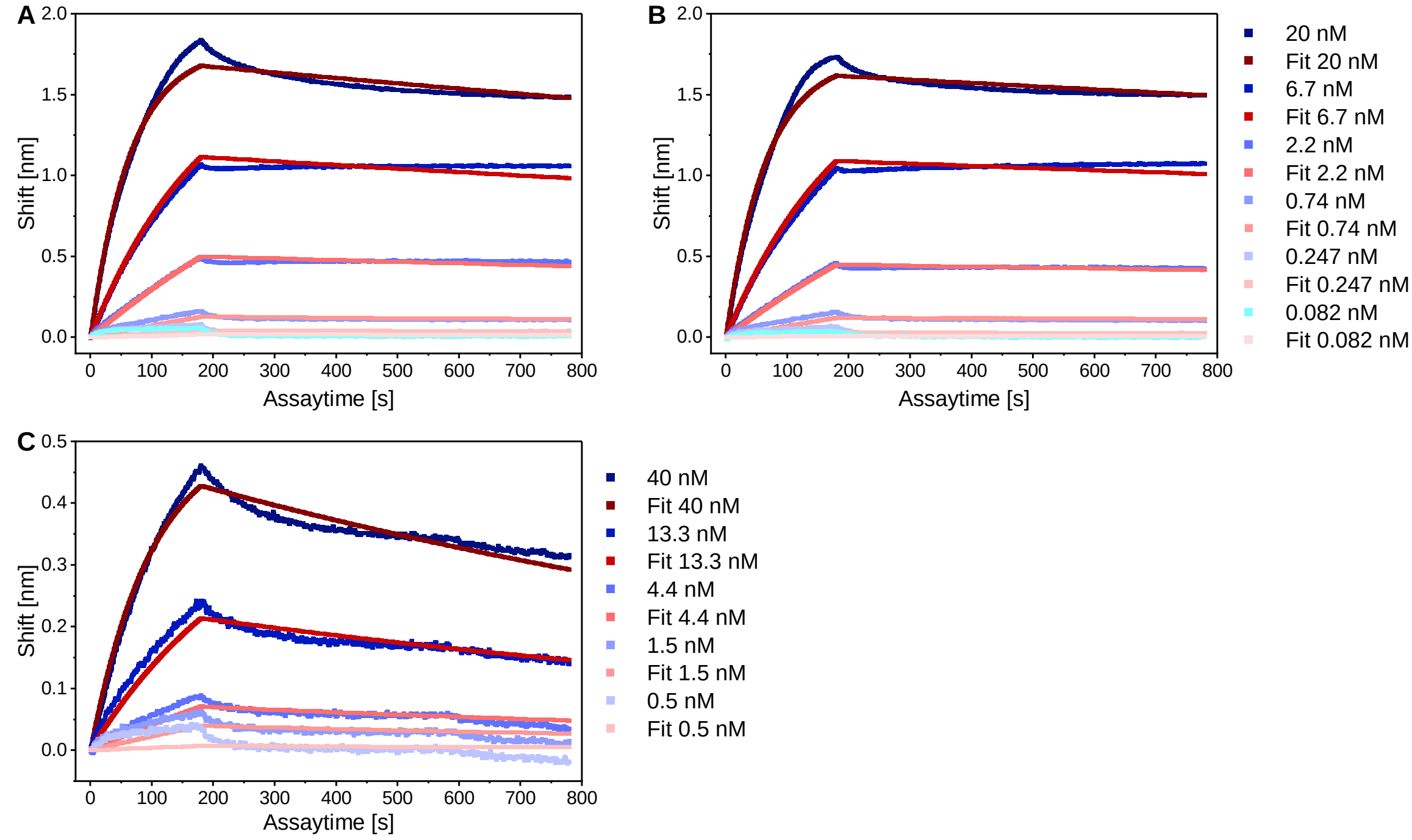

*Figure 1: Kinetic analysis of biotinylated scFv and corresponding antigens measured by BLI. (a): SH2424-B1 + N protein, (b): SH2424-H9 + N protein, (c): STE90-C11 + RBD25, (scFv + antigen). Sensorgrams (blue lines) show the association (180 s) and dissociation (600 s) phases of a 1:3 dilution of antigens (20 nM to 0.08 nM / 40 nM to 0.5 nM) to immobilized scFv. Global fitting to a 1:1 binding model with $R_{max}$ unlinked is shown in red lines. All measurements were carried out at 30 °C in kinetic buffer (TBS, 0.1 % PVP, 0.02 % Tween-20) under constant shaking at 1000 rpm. Calculated kinetic parameters (Kon, $K_{off}$, $K_{D,app}$, $K_D$) are summarized in*

*Table 1: Kinetic parameters of biotinylated scFv determined by BLI. Data was obtained by global fitting to a 1:1 binding model with $R_{max}$ unlinked. *Apparent equilibrium constant ($K_{D,app}$) determined using a dimeric antigen (N protein), reflecting potential avidity affects.*

| scFv + antigen | $K_{on}$ ($M^{-1}s^{-1}$) | $K_{off}$ ($s^{-1}$) | $K_{D,\,app}$ (M)* | $K_D$ (M) | $R^2$ |
|---|---|---|---|---|---|
| SH2424-B1 + N protein | $7.65 \times 10^5$ | $2.09 \times 10^{-4}$ | $2.73 \times 10^{-10}$ | - | 99.7 |
| SH2424-H9 + N protein | $7.56 \times 10^5$ | $1.28 \times 10^{-4}$ | $1.7 \times 10^{-10}$ | - | 99.7 |
| STE90-C11 + RBD25 | $2.35 \times 10^5$ | $6.35 \times 10^{-4}$ | - | $2.7 \times 10^{-9}$ | 98.98 |

## Effects of Antibody Functionalization on MNP

The functionalization of magnetic nanoparticles is a fundamental prerequisite for performing homogeneous MIA. However, the functionalization itself already increases the hydrodynamic diameter $d_h$ of the MNP. To maximize the signal change induced by analytes bound to the particle surface, the antibody-based functionalization should therefore increase the hydrodynamic size of the MNP as little as possible. Thus, it makes sense to investigate functionalization using only the scFv fragments instead of the entire IgG, since the scFv is significantly smaller. To determine the antibody concentration required for effective functionalization, the Antibody Functionalization Series described in Methods and Materials section is evaluated in the following. In Figure 2 the differences between BNF-D80 particles functionalized with biotinylated IgG (STE90-C11-IgG) and scFv (STE90-C11-scFv) are presented. The samples were analyzed using MPS as well as ACS for selected samples. When biotinylated IgG is added to the streptavidin-coated BNF-D80 particles, the $HR_{53}$ shows a significant decrease at a BBS occupancy of approximately 20 % (see Figure 2a). Subsequently, the ratio rises again with higher BBS occupancy and approaches a steady but smaller $HR_{53}$ compared to the initial value. The ACS measurement confirms this behavior, as the peak of the normalized imaginary part of the AC susceptibility $\chi''_{norm}$ initially shifts to frequencies below 10 Hz. The lowest peak frequency of 5 Hz is reached at a BBS occupancy of 20 % (see Figure 2b). Subsequently, the peak is observed to shift back toward higher frequencies as the BBS occupancy increases, with the shift gradually levelling off at 100 % occupancy. This property is unsuitable for sensitive functionalization, as obviously cross-linking is mediated by IgG. In cluster-based assays, however, cross-linking should be reserved for the analyte, as it has been shown to cause a significant change in the measurement signal like $HR_{53}$. In contrast, the MPS measurements on the scFv-functionalized MNP show a monotonically, nearly linear, decreasing trend in $HR_{53}$ over BBS occupancy (see Figure 2c) indicating no cluster formation. At a BBS occupancy at around 160 %, $HR_{53}$ approaches saturation at $HR_{53,sat}$, suggesting that the available binding sites have largely bound scFv antibodies. Theoretically, saturation should be reached at 100 % BBS occupancy. However, deviations may arise because particles may not fully match the datasheet specifications, leading to errors in the nominal occupancy calculation. Incomplete biotinylation of the scFv can provide a further source of deviation by preventing some scFv from binding in the functionalization process. Moreover, it should be noted that the decrease in $HR_{53}$ is small compared to that observed for IgG, suggesting a minor increase in the hydrodynamic diameter. These findings are confirmed by ACS measurement, which shows only a slight shift in peak frequency (see Figure 2d). Starting at 380 Hz the peak frequency reduces to 294 Hz at saturation, which corresponds to an increase in hydrodynamic diameter from 83.1 nm to 90.5 nm. The concentration-dependent curves obtained for SH2424-B1-scFv and SH2424-H9-scFv antibodies exhibit similar behavior (see Figure S1, Supporting Information).

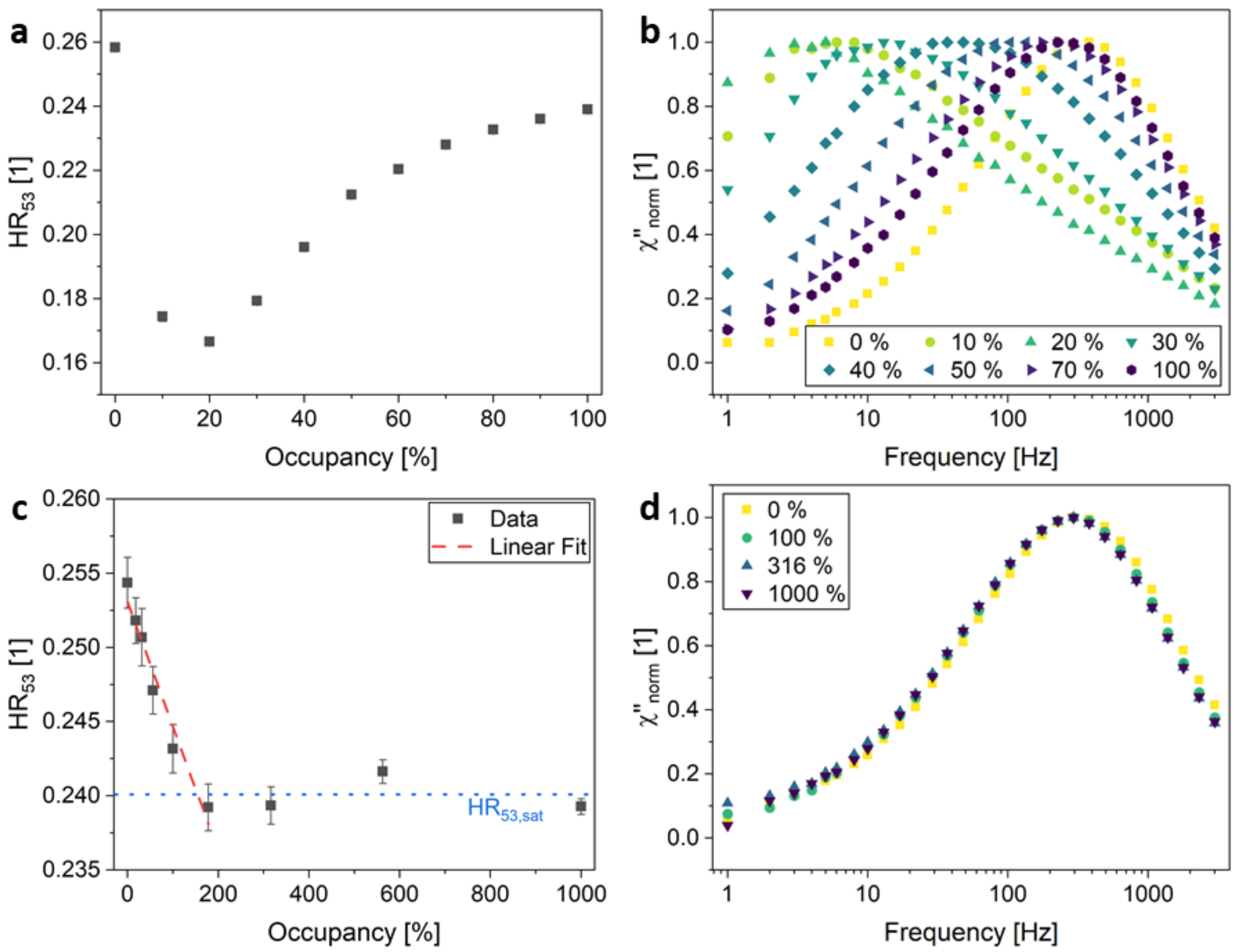

*Figure 2: Comparison of BNF-D80 particle functionalization with biotinylated IgG (STE90-C1-IgG) and scFv (STE90-C1-scFv) for different nominal biotin binding site (BBS) occupancies. The MPS and ACS measurements are represented by the harmonic ratio $HR_{53}$ and the normalized imaginary part of the AC susceptibility, $\chi''_{norm}$, respectively. For IgG functionalization, $HR_{53}$ and $\chi''_{norm}$ are shown in (a) and (b), respectively, whereas the corresponding measurements for scFv functionalization are shown in (c) and (d). The dashed line in (c) represents a linear fit. The $HR_{53,sat}$ indicator (dotted line) is calculated as the average of the three data points at the highest BBS occupancies.*

Consequently, it can be assumed that, the IgG antibodies likely contain several biotin molecules per antibody, allowing each MNP to form cross-links with the multiple possible BBS per MNP. This is consistent with the manufacturer's manual for the biotinylation kit, which reports that each IgG molecule contains 4 to 6 biotin molecules, and is therefore a drawback of random biotinylation. The scFv, with their monotonous behavior, indicate that they contain only one biotin molecule per antibody. Therefore, they are suitable for use in our MIA, and not merely because of their smaller size. Even though the MNP BBS reaches saturation at approximately the calculated antibody concentration, the samples should be oversaturated to ensure that all particles actually bind specifically to the analytes and can participate in the assay. The resulting necessary wash step, which is already recommended in [Kahmann2024] for higher assay sensitivity, removes the unbound scFv antibodies. This has an unavoidable impact on the particles, which is minimized as much as possible by the selected wash settings (see Figure S2, Supporting Information).

## Assay Response due to Cluster Formation

The effectiveness of the STE90-C1-scFv functionalized MNP for our homogeneous MIA is demonstrated below by detecting SARS-CoV-2 VLP. Here, the Cluster Formation Series described in Methods and Materials is analyzed. The ACS measurement in Figure 3a shows a distinct peak in $\chi''$ below 1 Hz at SARS-CoV-2 VLP concentrations above 2.61 pM. When converted to size, this results in clusters with a hydrodynamic diameter of more than 600 nm. For lower concentrations, the single particles peak at 294 Hz is dominant. Using MPS to measure the samples, a significant decrease is observed in $HR_{53}$ (see Figure 3b), which also suggests an increase in $d_h$. In comparison, the control with Hanta VLP remains stable across the entire concentration range, with a constant $HR_{53}$ value and consistent $\chi''$ curves (see Figure S3, Supporting Information). This confirmed that MIA could be successfully implemented with these materials. It is notable, however, that with MPS the lower VLP concentrations of 0.66 pM and 0.33 pM can no longer reliably resolved in $HR_{53}$ (value similar to that of the control without a target), while in $\chi''$ a prominent shoulder towards low frequencies froms. This change in relaxation behavior, which can be interpreted as a change in the MNP size distribution, should also be detectable using the MPS principle. For further investigation the 0 pM control and 0.13 pM SARS-CoV-2 VLP samples are subjected to field-dependent ACS measurements up to 5 mT/$\mu_0$, so that the susceptibility can be examined at field strengths that are closer to the 15 mT/$\mu_0$ used during the immunoMPS measurements. The results are presented in Figure 4. Figure 4a depicts the normalized imaginary part of the AC susceptibility $\chi''_{norm}$ under regular small-field excitation of 0.2 mT/$\mu_0$ as well as during large-field excitation of 5.0 mT/$\mu_0$. The normalized plot is particularly useful here for highlighting differences in the distribution and eliminating the influence of slight variations in MNP concentration. It is clearly evident that the peak frequency of both samples shifts upward, as would be expected based on the field-dependent Brownian relaxation time constant. At the same time, however, both curves converge when excited by 5.0 mT/$\mu_0$, which corresponds to an alignment of the distribution of the particles' effective relaxation times. Figure 4b illustrates this trend by showing the difference $\Delta\chi''_{norm}$ between the two samples. As the field strength increases, this difference generally decreases, and its maximum shifts to higher frequencies. For MPS measurements, this equalization of the effective relaxation times is problematic because it also causes the contributions to the individual harmonics, including the third and the fifth harmonic, to become equal in both samples, so that no change is detectable in the $HR_{53}$ ratio if the excitation field strength and frequency are chosen poorly. At low VLP concentrations, it is not expected that large clusters with many cross-links will form. Rather, it is reasonable to assume, due to the excess of MNP, small cluster formations in which the MNP primarily attach themselves around the VLP of similar size, provided that this is sterically feasible. One possible explanation is that the MNP interact with one another, thereby increasing the effective magnetic moment. Since this is a key parameter in the field-dependent Brownian relaxation time constant, these particles relax more quickly at higher field strengths, causing the $\chi''_{norm}$ curves to converge again at higher field strengths. However, the available data do not allow for a definitive conclusion regarding this finding, which is why further experimental and theoretical investigations are necessary. Furthermore, this behaviour in MPS is not observed in all cluster-based MIA, but occurs sporadically (see e.g. [Kahmann2024]), which suggests the influence of external factors in the implementation of the assays.

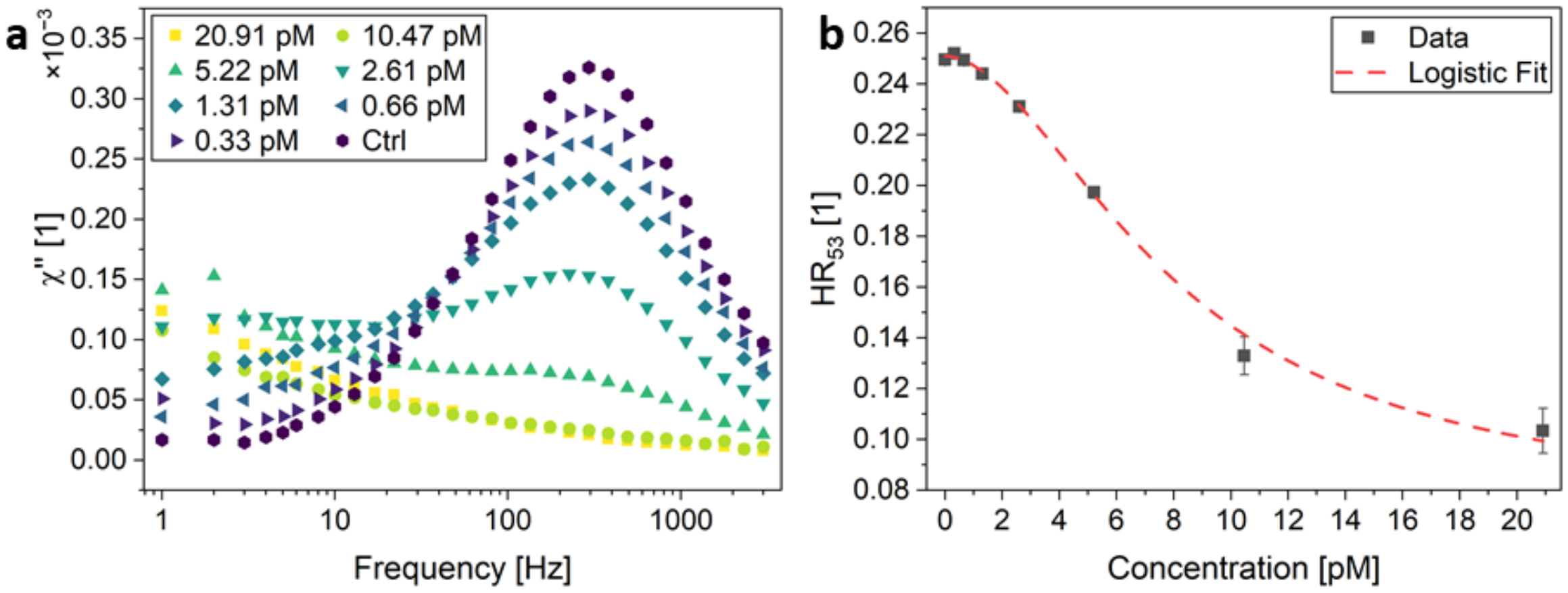

*Figure 3: Results of the assay characterization using MNP functionalized with STE90-C1-scFv and varying concentrations of SARS-CoV-2 VLP. The ACS measurement in (a) shows the imaginary part of the ac susceptibility $\chi''$, while the MPS measurement in (b) presents the harmonic ratio $HR_{53}$. The peak shift in $\chi''$ and the concentration-dependent change in $HR_{53}$ are illustrated for increasing VLP concentrations. The dashed line in (b) represents a logistic fit.*

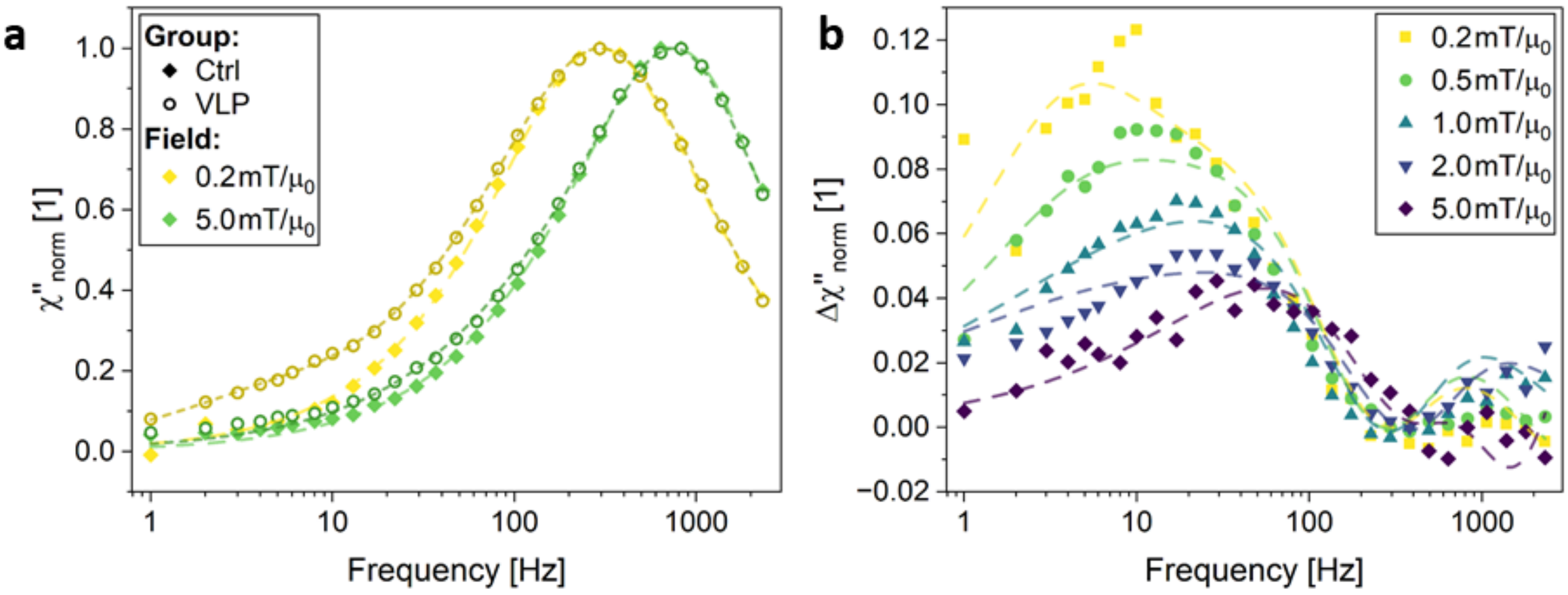

*Figure 4: Further investigation of the 0 pM control and 0.33 pM SARS-CoV-2 VLP samples using field-dependent ACS. Shown is (a) the normalized imaginary part of the susceptibility $\chi''_{norm}$ of both samples under standard small-field excitation of 0.2 mT/$\mu_0$ and a stronger field of 5.0 mT/$\mu_0$, as well as (b) the difference between both curves $\Delta\chi''_{norm}$ for several field strengths. The dotted lines represent phenomenological fits as guides to the eyes.*

## Influence of Particle Concentration on Assay Sensitivity

The MNP concentration is a factor that directly influences the MIA and its cross-linking, because for a given target concentrations, the ratios and thus the binding probabilities change. This is accompanied by changes in the magnetic signal response, so that changes in the concentration-independent harmonic ration $HR_{53}$ of an MPS measurement can be expected solely as a result of the altered cross-linking. Consequently, to optimize the assays, it is very important to investigate their sensitivity by varying the MNP concentration. To this end, the samples from the Assay Response Series described in Methods and Materials are analyzed. Samples below 0.2 nM MNP are measured exclusively using the MPS setup, since the signal strength is no longer sufficient to produce meaningful ACS results. The MPS measurement results for the SARS-CoV-2 VLP experiments are shown in Figure 5a,b, while the assays for the N protein are presented in Figure 5c,d. The corresponding ACS curves for the 0.2 nM MNP samples can be viewed in Figure S5 in the Supporting Information.

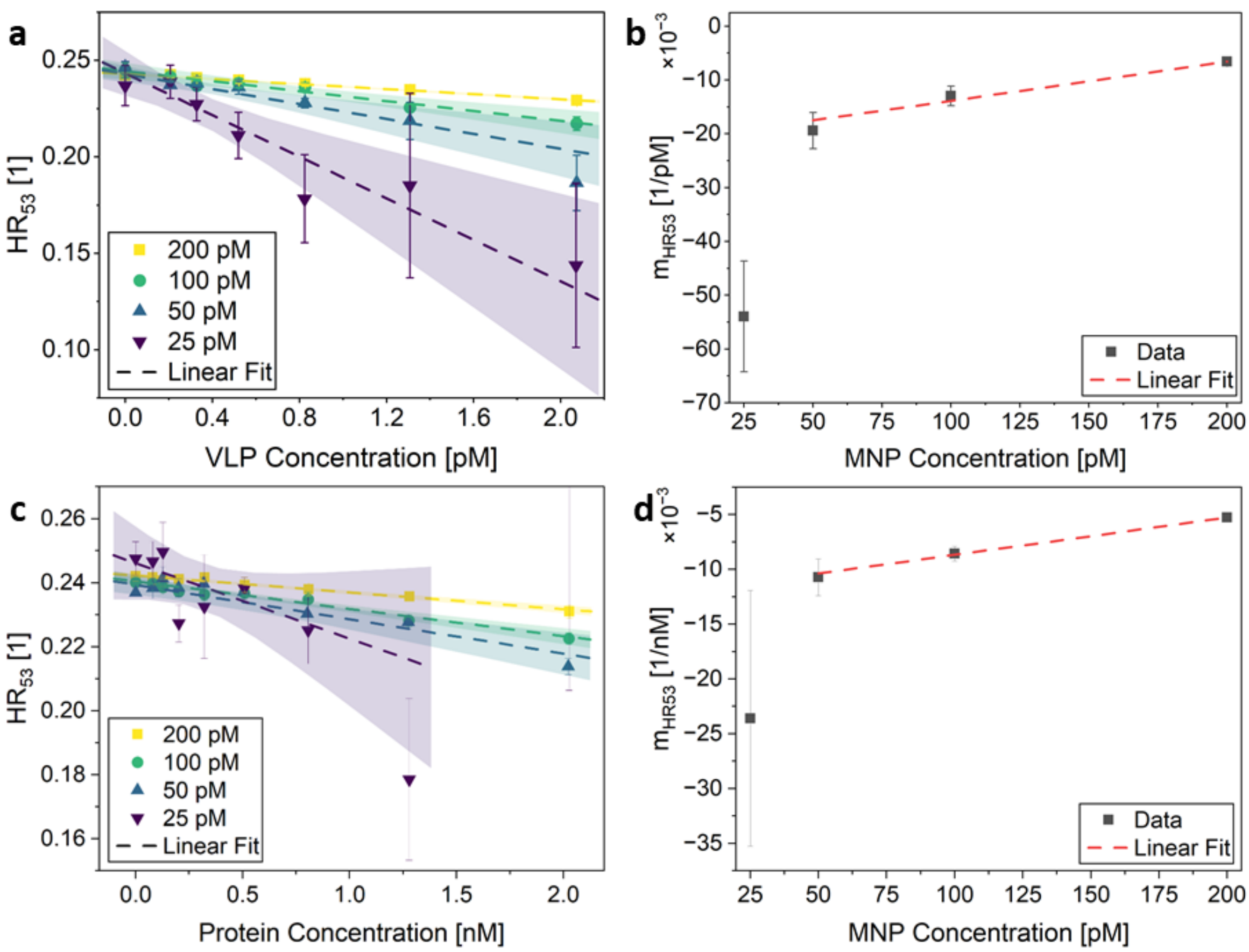

*Figure 5: Dependence of the harmonic ratio $HR_{53}$ and assay sensitivity on analyte concentration at different MNP concentrations for SARS-CoV-2 VLP and N protein. For SARS-CoV-2 VLP, (a) shows $HR_{53}$ as a function of analyte concentration and (b) the corresponding slopes $m_{HR53}$ of the linear fits as a function of MNP concentration. The respective data for N protein are shown in (c) and (d). $HR_{53}$ measurements were performed at MNP concentrations of 0.2 nM, 0.1 nM, 0.05 nM, and 0.025 nM, with the displayed analyte concentration range restricted to the approximately linear regime of the assay response. Dashed lines represent linear fits to the individual data sets, with 95 % confidence intervals shown as transparent bands. Error bars in (b) and (d) indicate the standard errors of the fitted slopes.*

For the SARS-CoV-VLP assays cluster formation is generally to be expected due to the several S proteins and therefore binding sites on the surface of the VLP as previously verified with ACS measurements (see Figure 3a) and confirmed for this experiment (see Figure S5, Supporting Information). In MPS measurements, Figure 5a shows that the $HR_{53}$ ratio decreases approximately linearly with SARS-CoV-2 VLP concentration below 2.1 pM. Only the 0.2 nM MNP samples exhibit a slight plateau at vanishing VLP concentrations before the $HR_{53}$ declines linearly. A linear fit is performed for all samples and indicated by the dashed lines for the sake of clarity. The slopes of the linear fits become progressively steeper as the MNP concentration is lowered. At the same time, the standard deviation of the individual data points becomes larger, since the signal-to-noise ratio (SNR) diminishes at lower MNP concentrations (see Figures S7 and S8, Supporting Information). Consequently, the 95 % confidence interval broadens, as indicated by the transparent band around the respective fit. The slope and the corresponding standard error of each linear fit are extracted and plotted separately in Figure 5b. The enhanced sensitivity at lower MNP concentrations, due to a larger absolute value of the slope $m_{HR53}$, is clearly evident. Furthermore, the data points indicate a sharp rise in sensitivity at lower MNP concentrations, following an exponential rather than a linear relationship. However, the standard error of the slopes, which increases severe due to the low SNR, makes it difficult to draw definitive conclusions. The overall trend is consistent with expectations, since larger clusters can form for the same VLP but lower MNP concentrations, just as they do when the VLP concentrations increases at a constant MNP concentration, thereby amplifying the drop in $HR_{53}$ (see Figure 3b).

The same analysis can be performed for the detection of N protein using our MIA. It should be noted that clusters also form during the detection of N proteins, which can be attributed to their dimer structure. This is illustrated by the ACS data for the 0.2 nM MNP samples (see Figure S5, Supporting Information), which shows the formation of a broad shoulder in the single-digit Hz range. Figure 5c examines the largely linear region of the $HR_{53}$ ratio from the MPS measurement, which is observed for concentrations below 2.2 nM of N protein. The recorded data points for the harmonic ratio $HR_{53}$ are fitted linearly again for better description, and the 95 % confidence intervals are indicated by the corresponding transparent bands in the background. As before, the slope of the linear fit becomes steeper as the MNP concentration is reduced. Compared to the VLP, however, the slope is less pronounced. It can even be found that, in the samples containing only 0.025 nM MNP, the measurement points fluctuate significantly making the analysis more difficult. The data point for the 2.1 nM N protein sample had to be excluded from the fit as an outlier in order to perform the fit properly. With only 0.025 nM of particles and approximately 107 antibodies per MNP, calculations show that at around 2.5 nM of N protein, every binding site is occupied. Therefore, the formation of large clusters and a shift from linear behavior toward saturation behavior can be expected in this range. Generally, this is accompanied by a strong reduction in the amplitudes of the individual harmonics (see Figures S7, S8, Supporting Information), resulting in a sharp deterioration in the SNR and thus large standard deviations. Figure 5d plots the slopes extracted from the linear fits and their standard errors against the MNP concentration. Excluding the data point at the lowest MNP concentration for a moment, the data suggest a fairly linear increase in the absolute value of the slope, and thus in the MIA's sensitivity, as the MNP concentration drops. However, a slight trend toward a more pronounced increase in sensitivity with decreasing MNP concentration can still be observed, particularly when the data point with the lowest MNP concentration is also included in the analysis.

A comparative analysis of the two assay variants reveals a 3000 up to 4000 times more sensitive VLP assay compared to the N protein assay in the linear detection range. In addition, VLP assays measure concentrations that are three orders of magnitude below the detectable N protein levels. This indicates not only the importance of cluster formation for a good measurement effect, but also and above all that the size of the analyte to be detected should be at least comparable to that of the magnetic nanoparticle. This is obvious, since the relative change in hydrodynamic size upon binding of a molecule is greater if the molecule itself is larger. Moreover, the large number of S proteins, and thus potential binding sites, on the surface of a VLP facilitates the formation of cross-links, which also contributes to this result. To distinguish the influence of these two effects on the assay performance further investigations are necessary.

## Detection Limit Evaluation

To assess the performance of our scFv-optimized assays, an estimate of the LOD is made. For this purpose, the samples form the Detection Limit Series in Methods and Materials containing only 0.05 nM and 0.025 nM MNP are considered, as these exhibit the highest sensitivity and thus promise the potentially best limit of detection, provided that the SNR is still sufficiently high. The evaluation is performed according to the procedure described in Methods and Materials under Detection Limit Determination, taking all three independent samples into account. Figure 6a and Figure 6b show the results of the MPS measurements in the LOD series for SARS-CoV-2 VLP and N proteins, respectively, at an MNP concentration of 0.05 nM. The mean value of the harmonic ratio $\overline{HR}_{53}$ shown here represents the mean of the three preparations, while the standard deviation refers to the variance of an individual measurement, so fully including the variability of multiple measurements and the variability of sample preparation. In addition, the negative control with the highest concentration of the analyte being tested is displayed for comparison in each case. In the first case, it consists the Hanta VLP, whereas in the second it compromises the S1 proteins. This clearly demonstrates that both assays exhibit high specificity, as the negative controls yield values comparable to those of the blank sample. The estimated LOD is determined from the point of intersection between the linear regression of the $HR_{53}$ versus analyte concentration and the detection threshold $\overline{HR}_{53,thr}$. For the SARS-CoV-2 VLP, this is approximately 490 fM, while for the N proteins it leads to 1.7 nM. This corresponds to a particle number density of $2.95{\cdot}10^{8}$ VLP/mL and $1.02{\cdot}10^{12}$ proteins/mL, respectively. The data for the samples with an MNP concentration of only 0.025 nM can be found in the Figures S10 in the Supporting Information. The analysis is performed in the same manner. For the SARS-CoV-2 VLP series, the limit of detection is 649 fM or $3.91{\cdot}10^{8}$ VLP/ml. This is higher than for the 0.05 nM MNP samples, which is attributable to the increased standard deviation in $HR_{53}$ due to the reduced SNR, even though the sensitivity itself increases (see Figure 5b). In the case of N proteins, no reliable linear range can be identified for the 0.025 nM MNP samples, and therefore no limit of detection is specified here. It should be noted that the reported detection limits are influenced by the previously determined concentrations of the stock solutions, which introduce an additional source of uncertainty.

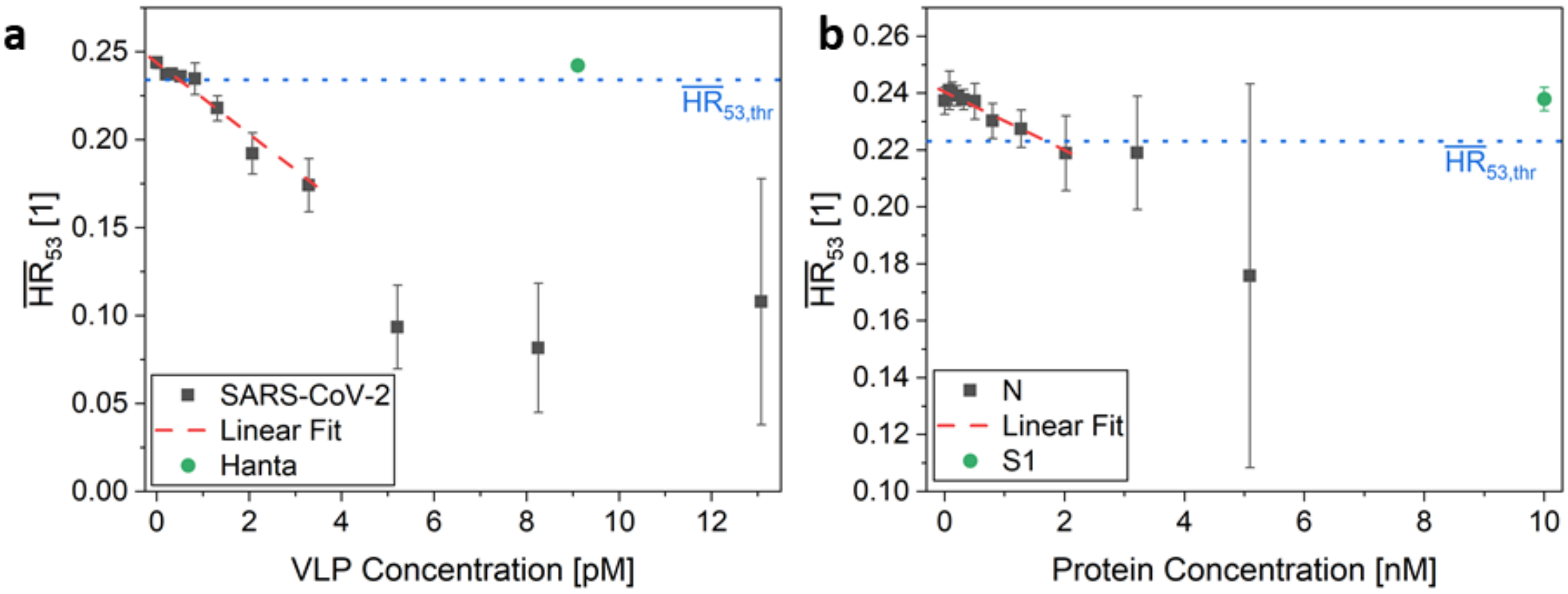

*Figure 6: Limit of detection (LOD) determined by MPS at an MNP concentration of 0.05 nM. The averaged harmonic ratio $\overline{HR}_{53}$ of the three independent preparations is plotted against the analyte concentration of (a) SARS-CoV-2 VLP and (b) N proteins. The standard deviation of each point refers to the variance of an individual measurement without repetitions. Furthermore, the negative controls at the highest analyte concentrations are plotted, which are (a) Hanta VLP and (b) S1 proteins. The LOD is determined by the point where the linear regression (dashed line) intersects the detection threshold (dotted line).*

In both cases, the detection limit for SARS-CoV-2 VLP represents a slight improvement on the estimate made in earlier studies using mimic viruses, for which a detection limit of 552 fM was determined [Wolgast2024]. In addition, this study takes the variability resulting from the different preparations in the LOD estimation into account indicating an even bigger progression. For the N protein, a LOD of 12.5 nM was reported in comparable studies [Wu2021]. The assays we have carried out therefore have an improved limit of detection. However, the previous analysis of the assay's sensitivities suggests that the use of smaller nanoparticles is particularly promising for the improved detection of N protein.

To contextualize the LOD achieved with our magnetic assays for SARS-CoV-2 VLP, we compare it with the established ELISA method. The corresponding results are shown in Figure 7, which plots the absorbance. The data were fitted using a logistic function, the blank sample shown consists of BSA + TMB + $H_2SO_4$. The sample containing Hanta VLP serve as negative control to verify the specificity of the assay. The data were analyzed in the same manner as the MPS data, using three samples with five measurements each, from which the mean and standard deviation for a future single measurement were determined.

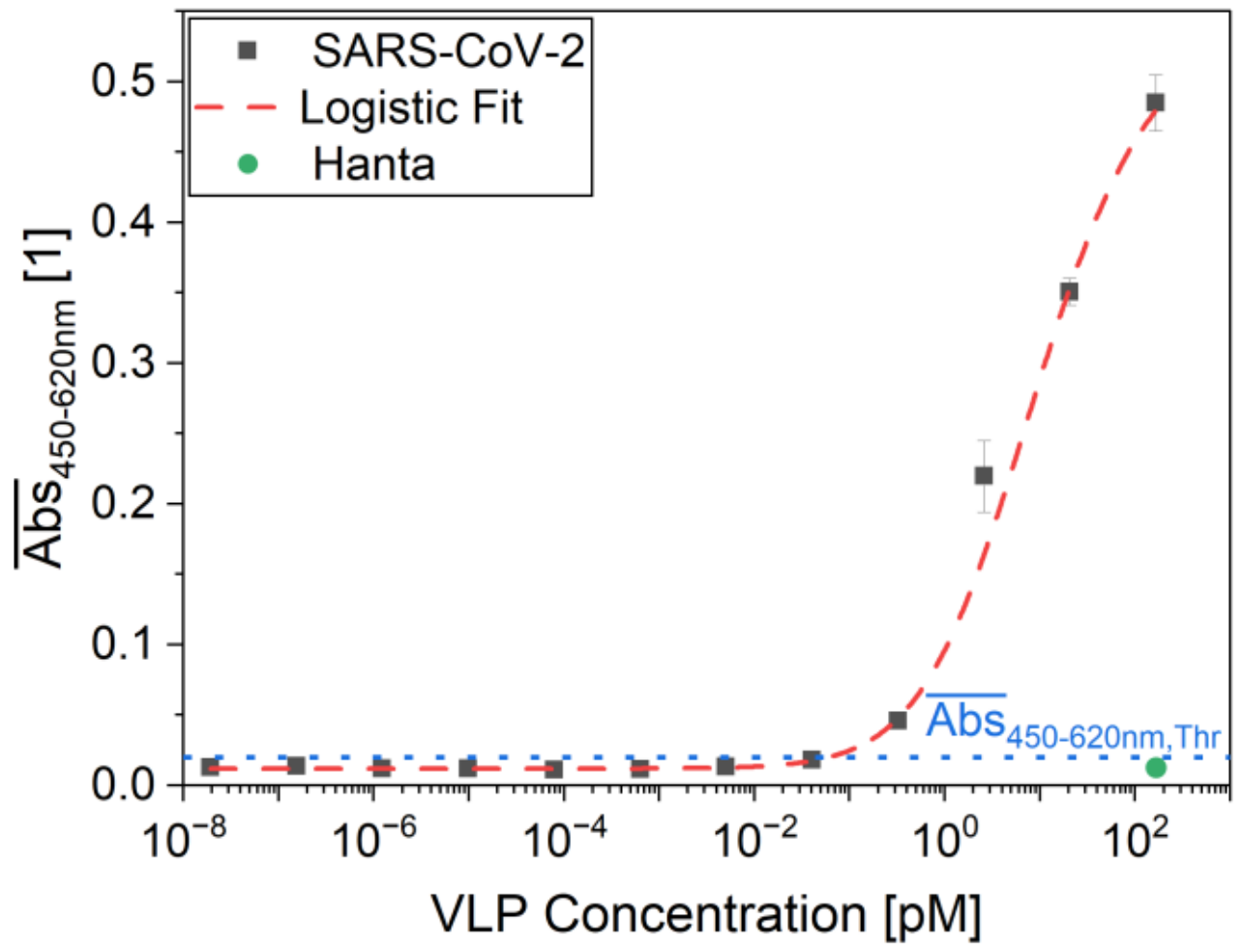

*Figure 7: Determination of the Limit of Detection in an ELISA for SARS-CoV-2 VLP. The averaged absorption $Abs_{450-620nm}$ of three independent preparations is plotted against the SARS-CoV-2 VLP concentration. The standard deviation of each point refers to the variance of an individual measurement without repetitions. Furthermore, the negative control consisting of highly concentrated Hanta VLP is shown. The LOD is determined by the point where the logistic regression (dashed line) intersects with the detection threshold $Abs_{450-620nm,Thr}$ (dotted line).*

The LOD is also estimated using the 3-sigma criterion and is found to 58 fM. Thus, the established method is still about a factor of 9 better, however, it is evident that magnetic assays can yield nearly comparable results and thereby expand the range of evaluation options.

# Conclusion

In the present study, we have developed magnetic immunoassays with scFv-functionalized BNF-D80 nanoparticles for detecting SASR-CoV-2 VLPs or Nucleocapsid proteins. The scFv antibodies against the S and the N protein were first tested using BLI to determine their functionality and their association and dissociation kinetics. This was followed by functionalizing the streptavidin-coated BNF-D80 particles with various concentrations of the biotinylated antibodies to investigate the loading and magnetic behaviour of the MNP. The difference between the full-length IgG and the scFv fragment was also examined, with the IgG subjected to random biotinylation and the scFv undergoing targeted biotinylation during production. It was demonstrated that, unlike the IgG, the scFv do not form parasitic cluster structures during MNP functionalization, meaning that there is at most one biotin per scFv. At the same time, the scFv enabled complete functionalization of the MNPs with only a minimal increase in hydrodynamic diameter of around 6 nm.

To demonstrate cluster formation in our MIA, a series of measurements was conducted using various SASR-CoV-2 VLP concentrations, which revealed both cluster formation in the ACS and a decrease in the harmonic ratio $HR_{53}$ in the MPS. At the same time, it was observed that, under certain conditions, cluster formation can lead to a reduction in the measurement effect at higher field strengths, such as those used in MPS. This makes it considerably more difficult to distinguish positive samples with low analyte concentrations from negative samples in $HR_{53}$. To overcome this effect in magnetic assays when analyzing results with MPS, the assay responses were examined at various MNP concentrations to increase sensitivity. SARS-CoV-2 VLPs and N proteins were used as analytes. The results showed that N proteins also form clusters, which, in the case of N proteins, is attributable to their dimeric structure. As expected, sensitivity increases as the number of MNPs decreases in both cases, because a relatively larger proportion of the MNP then bind to the analyte, making it easier to measure. Furthermore, the results showed that SASR-CoV-2 VLPs exhibit a sensitivity 3000 to 4000 times higher than that of the N proteins within the linear detection range, and that concentrations three orders of magnitude lower can be detected using the VLPs. The quantitative detection limits for these two analytes in our magnetic assays were determined by extending the analyte concentration series to three independent samples. For the particle concentration, only those samples that achieved the previously determined highest sensitivity were considered. The LOD was estimated using the 3-sigma criterion, with the standard deviation accounting for both the variability of multiple measurements and the variability of sample preparation. Detection limits of 490 fM ($2.95 \cdot 10^{-8}$ VLP/ml) SARS-CoV-2 VLP and 1.7 nM ($1.02 \cdot 10^{12}$ proteins/mL) were determined, representing an improvement over previous studies. At the same time, the negative controls demonstrated the assay's high specificity. Additionally, it became apparent that samples with an even lower MNP concentration exhibited a worse detection limit, which is attributable to a poorer signal-to-noise ratio and thus negates the advantage of higher sensitivity. Choosing the correct amount of particles is therefore a critical task.

## Supporting Information

Available upon request.

## Acknowledgements

This work is funded by the Deutsche Forschungsgemeinschaft (DFG, German Research Foundation) – 503667663. The authors gratefully acknowledge the support provided by Braunschweig International Graduate School of Metrology B-IGSM.